\documentclass[aps,prl,twocolumn,showpacs,10pt,superscriptaddress]{revtex4-1}

\usepackage{amssymb}
\usepackage{graphicx}
\usepackage{amsfonts}
\usepackage{amsmath}
\usepackage[usenames]{color}

\begin{document}

\title{Optimization of STIRAP-based state transfer under dissipation}
\author{Ying-Dan Wang}
\email{yingdan.wang@itp.ac.cn}
\affiliation{Institute of Theoretical Physics, Chinese Academy of Sciences, Beijing 100190, China}

\author{Xiao-Bo Yan}
\affiliation{Institute of Theoretical Physics, Chinese Academy of Sciences, Beijing 100190, China}

\author{Stefano Chesi}
\email{stefano.chesi@csrc.ac.cn}
\affiliation{Beijing Computational Science Research Center, Beijing 100084, China}

\date{\today}

\begin{abstract}
Using a perturbative treatment, we quantify the influence of non-adiabatic leakage and system dissipation on the transfer fidelity of a stimulated Raman adiabatic passage (STIRAP) process. We find that, optimizing transfer time rather than coupling profiles, leads to a significant improvement of the transfer fidelity. The upper bound of the fidelity has been found as a simple analytical function of system cooperativities. We also provide a systematic approach to reach this upper bound efficiently.

\end{abstract}

\maketitle



\emph{Introduction.-} State transfer, where an arbitrary quantum state is transmitted from a source to a target system, is a fundamental task in quantum state engineering. While sometimes it is possible to directly couple quantum systems (even of different nature), in many circumstances the transfer must be mediated by a third party (a quantum bus), whose coherence properties play a crucial role for the transfer process. To mitigate the effect of coupler decoherence, a particularly interesting class of indirect transfer protocols is based on the idea of stimulated Raman adiabatic passage (STIRAP), as it allows to perform the state transfer with negligible excitation on the quantum bus. 

STIRAP was developed more than 2 decades ago for population transfer in $\Lambda $-type atoms \cite{Gaubatz1990}. Due to the advantages of being robust, simple, and efficient, this approach, as well as its theoretical extensitons~\cite{Bergmann1998, Vitanov2001, Petr2007, Sola1999, Grigoryan2009, Torosov2013, Stefano2015},  have found application in a variety of physical systems and for many different tasks (see, e.g., Refs.~\cite{Weitz1994, Kulin1997, Duan2001, Kuhn2002, Oh2013, Mucke2013, Petrosyan2013, Drummond2002, Ling2004,Sun2003, Klein2007, Lang2008, Moller2008, Wang2012, Wang2012NJP, Webster2013, Koh2013, Golter2014, Trautmann2015, Dupont2015}). One of the most relevant applications of STIRAP is when a discrete bosonic mode (e.g., of a micromechanical resonator, microwave cavity, or optical cavity) mediates the state transfer between two qubits (either of different types such as the hybrid systems of superconducting qubits and NV centers, or distant qubits of the same type). Furthermore, one or both of the qubits can be replaced by a cavity mode (see, e.g.,~\cite{Andrews2014,Tabuchi2015}). 

While the optimization of STIRAP for atomic population transfer was performed in a decoherence-free subsystem~\cite{Vasilev2009}, with the objective of reducing the non-adiabatic leakage and of minimizing the pulse area or duration, the interplay between decoherence and non-adiabatic transitions is a crucial aspect for many current applications. Different from atomic systems, most solid-state quantum devices suffer significant dissipation. Hence, the prolonged operation time of STIRAP (required by adiabaticity) becomes a severe drawback: even if the source and target are relatively weakly coupled to the environment, the accumulation of errors over a long transfer time could still significantly degrade the transfer fidelity. Thus, in the presence of dissipation, optimization of the coupling profiles is essential to reach a trade-off between the non-adiabatic leakage and system dissipation, and to guarantee a high-fidelity state transfer.

To the best of our knowledge, a general understanding of this trade-off, as well as the resulting fundamental limitations, is missing in the literature. Some previous works considering the effect of dissipation on STIRAP have only included the decoherence of the intermediate level~\cite{Ivanov2005,Yuan2012,Scala2010}, often with a phenomenological approach  \cite{Ivanov2005,Goto2008}. In other studies, the dissipation is numerically simulated (see, e.g.,~\cite{Greentree2004,Vogt2012,Hou2013,Dupont2015}). In order to provide physical insight and determine the power of STIRAP in the most realistic scenario - and especially to clarify the crucial interplay by the two types of dissipation mechanisms - we pursue here an analytical treatment, with full consideration of the system dissipation. We solve the master equation with a perturbative approach which is accurate in the desired high-fidelity regime. Although it is usually believed that the coupling profiles play an important role in optimizing the protocol, we quantify their influence and find it is relatively minor. Instead, the upper bound of the STIRAP fidelity critically depends on the transfer time and it is decided by a simple function of the cooperativities. We also provide a systematic optimization procedure to reach this upper bound efficiently. Our results can be applied to a variety of physical implementations such as optomechanics, circuit QED, and hybrid systems~\cite{Aspelmeyer2014,Xiang2013}.

\begin{figure}
\begin{centering}
\includegraphics[width=0.48\textwidth]{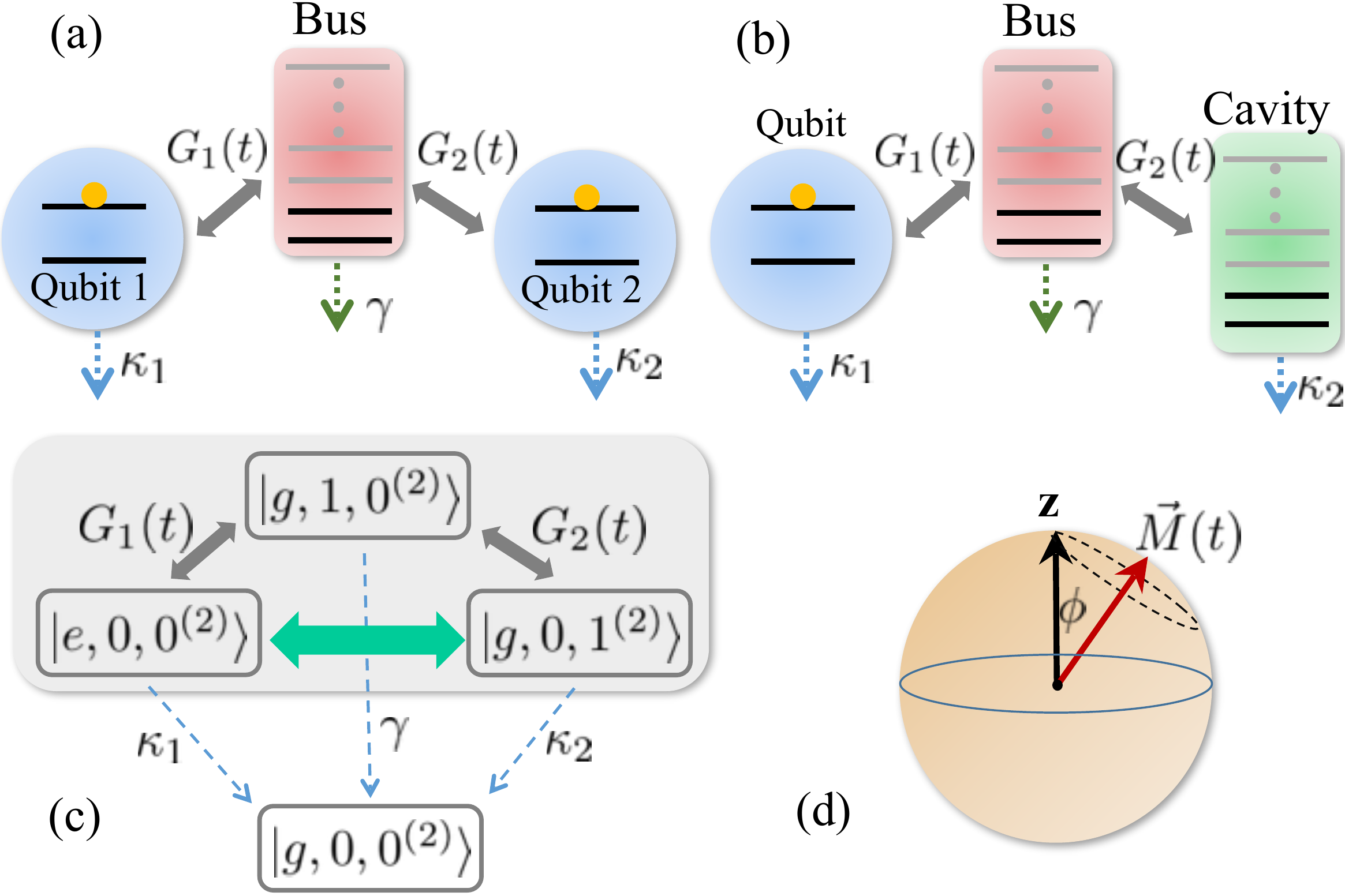} 
\par\end{centering}
\caption{(a) The schematics of transfer setup between two qubits coupled via a common quantum bus. (b) The schematics of transfer setup between a qubit and cavity. We assume the higher energy levels (grey) are not populated. (c) The energy diagram of the STIRAP based state transfer. The part in shadow corresponds to a $\Lambda$ atom in the original STIRAP protocol. The green arrow indicates the desired transfer of the excited state of a qubit and this is realized by tuning coupling $G_1(t)$ and $G_2(t)$ as STIRAP. The dissipation brings all the excited states to the ground state. (d) The unitary time evolution of the STIRAP can be represented by the rotation of a spin-1 vector.}
\label{schematics}
\end{figure}

\emph{System and dynamics.-} We consider the setup schematically illustrated in Fig.~\ref{schematics}(a), where two qubits interact resonantly with a common bosonic mode. Using the rotating wave approximation, the Hamiltonian in the interaction picture is $H^{\mathrm{I}}=\sum_{i=1,2} G_{i}\left( t\right) \left( a^{\dagger }\sigma^{(i)} _{-}+a\sigma^{(i)}_{+}\right) $, with $G_i(t)$ the tunable coupling strengths, $a$ the bosonic annihilation operator of the bus, and $\sigma _{\pm }=\sigma _{x}\pm i\sigma _{y}$. Such a Jaynes-Cummings Hamiltonian has been realized in cavity QED and various circuit QED architectures.  Considering dissipation, the bus and qubits decay  to their ground states with rates $\gamma $ and $\kappa _{i}$, respectively. The system dynamics is described by the master equation$\ d{\rho }^{\mathrm{I}}/dt=-i[H^{\mathrm{I}
},\rho ^{\mathrm{I}}]+\mathcal{L}\rho ^{\mathrm{I}}$, with $\mathcal{L=L}_{m}+\sum_{i}\mathcal{L}_{\mathrm{q}}^{\left( i\right) }$ and Lindblad dissipators $\mathcal{L}_{\mathrm{m}} = \gamma \mathcal D [a] $ and $\mathcal{L}_{\mathrm{q}}^{\left( i\right) } = \kappa _{i}   \mathcal D [\sigma_-^{(i)}] $, with $\mathcal D [A] \rho = A\rho
A^{\dagger }-\frac12\left\{ A^{\dagger }A,\rho \right\}$. Our discussion can also be easily extended by including pure dephasing terms $ \frac12 \gamma _{\varphi }^{(i)}  \mathcal D [\sigma_z^{(i)}] $~\cite{EPAPS}. This description of the dynamics is commonly adopted in cavity QED, superconducting qubits, optomechanical systems, and might capture the main features of other types of qubits as well, although their decoherence dynamics in some cases can be more complex (see, e.g.,~\cite{Stefano2016}).

The general goal is to
transfer an arbitrary state $|\psi \rangle =c_{g}\left\vert g\right\rangle
+c_{e}\left\vert e\right\rangle $ from qubit 1 to qubit 2. Considering a
sufficiently low environmental temperature (to allow at most one
excitation), the state transfer protocol is confined in a 4-level subspace formed by
$\left\vert 1\right\rangle =\left\vert e^{\left( 1\right) },0,g^{\left(
2\right) }\right\rangle$, $\left\vert 2\right\rangle =\left\vert
g^{\left( 1\right) },1,g^{\left( 2\right) }\right\rangle$, $\left\vert 3\right\rangle =\left\vert g^{\left( 1\right) },0,e^{\left(
2\right) }\right\rangle$, $\left\vert 4\right\rangle =\left\vert
g^{\left( 1\right) },0,g^{\left( 2\right) }\right\rangle$. It is clear that in this low-excitation limit, each party can be either a qubit or  a bosonic mode, i.e., the discussion below is also applicable to the case of two qubits coupled through a third qubit~\cite{Braakman2013,Oh2013} or a solid-state qubit coupled to an optical cavity through a mechanical oscillator~\cite{Stannigel2010,Stannigel2011} (as illustrated in Fig.~\ref{schematics}(b)).

An ideal transfer corresponds to $c_{g}\left\vert 4\right\rangle
+c_{e}\left\vert 1\right\rangle \rightarrow c_{g}\left\vert 4\right\rangle
+c_{e}\left\vert 3\right\rangle $. At zero temperature, state $|4\rangle$ is stable and all the loss of fidelity is due to the transfer of the excited state. Hence in the following we only discuss $c_{g}=0$, which is the most demanding case and it is equivalent to a pure STIRAP process (marked by the blue shadow in Fig.~\ref{schematics}(c)).
The idea of STIRAP is to adiabatically tune  $G_{1}(t)$ from
zero to a finite value, while $G_{2}(t)$ is tuned from a finite value
to zero such that the system evolves from $|1\rangle$ to $|3\rangle$. In the whole process, the system adiabatically follows the instantaneous eigenstate (dark state):
\begin{equation}
\left\vert \tilde{2}(t)\right\rangle  =-\cos \theta (t)\left\vert
1\right\rangle +\sin \theta (t)\left\vert 3\right\rangle , \label{dark}
\end{equation}
with $\tan \theta (t)=G_{1}(t)/G_{2}(t)$ and 
\begin{equation}
\theta (0)=0,\quad \theta (t_{\mathrm{f}})=\pi /2, \label{boundary}
\end{equation}
where $t_\mathrm{f}$ is the final operation time and it determines the overall speed of the transfer.

The density matrix $\tilde{%
\rho}=U^{\dag }\left( t\right) \rho ^{I}U\left( t\right) $ (where $U\left(
t\right) =\sum_{k}|\tilde{k}\left( t\right) \rangle \langle k|$ and $\{ |\tilde{k}(t)\rangle \} $ are the instantaneous eigenstates of $H^{\mathrm{I}}$ \cite{EPAPS}) satisfies:
\begin{equation}
\frac{d\tilde{\rho}(t)}{dt}=-i[\tilde{H}(t),\tilde{\rho}(t)]+\frac{\dot{
\theta}(t)}{\sqrt{2}}[\mu ,\tilde{\rho}(t)]+\mathcal{\tilde{L}}\tilde{\rho}(t),  \label{eqcomplete}
\end{equation}%
where $\tilde{H}(t)=G(t)(|{1}\rangle \langle {1}|-|{3}
\rangle \langle {3}|)$, $\mu =(|{2}\rangle \langle {1}|+|{3}\rangle \langle {2}|-\mathrm{h.c.})$, and $G\left( t\right) =%
\sqrt{G_{1}^{2}\left( t\right) +G_{2}^{2}\left( t\right) }$. The transformed
dissipator is defined by $\mathcal{\tilde{L}}\tilde{\rho}=U^{\dag }\mathcal{L}(U\tilde{\rho}U^{\dag })U$ and its matrix form can be found in a straightforward way~\cite{EPAPS}.
The last two terms in Eq. (\ref{eqcomplete}) corrupt the desired transfer process: the first one represents the
non-adiabatic leakage out of $\left\vert \tilde{2}(t)\right\rangle $, which dominates for a fast-changing pulse; while the second term, i.e., the environment dissipation, dominates for a slow-changing pulse.  How to reach an optimal trade-off between the two effects will be the central issue in the following sections. Before moving to that discussion, it is worth mentioning that the unitary part of Eq.~(\ref{eqcomplete}) describes a fast spin precession around $\vec M(t)=G(t) \hat{\bf e}_{z}+\dot{\theta}%
(t)\hat{\bf e}_{y}$, using a spin-1 language~\cite{EPAPS}. A perfect STIRAP state transfer can be realized when $G$, $\dot{\theta}$ are constant and the transfer time is a multiple of the precession period (see Fig.~\ref{schematics}(d)). In the absence of dissipation, there are also various other strategies to realize a perfect state transfer~\cite{EPAPS}.

\emph{Perturbative treatment.-} A successful transfer requires that the adiabatic dynamics plays a dominant role.
Hence the effect of the non-adiabatic leakage and dissipation in Eq.~(\ref{eqcomplete}) can be treated perturbatively. To do this, the density matrix is expanded as $\tilde{\rho}(t)=\tilde{\rho}^{(0)}(t)+%
\tilde{\rho}^{(1)}(t)+\tilde{\rho}^{(2)}(t)+\cdots $, which yields a corresponding expansion for the transfer fidelity
$F=\sum_{k=0}^{\infty }F^{(k)}=\sum_{k=0}^{\infty }\tilde{\rho}_{22}^{\left(k\right) }\left( t_{\mathrm{f}}\right) $ (using the boundary condition Eq.~(\ref{boundary})). The lowest-order result is $F^{\left( 0\right) }=1$ and the higher-order contributions are obtained by iterative solution of the equation for $\tilde{\rho}^{(k)}(t)$. Calculation details and the expressions including dephasing can be found in Ref.~\cite{EPAPS} and we report here only the simplified final results without dephasing. The first-order correction reads:
\begin{equation}\label{F1}
F^{\left( 1\right) }\mathcal{=-}\int_{0}^{t_{\mathrm{f}}}\left( \kappa
_{1}\cos ^{2}\theta (\tau)+\kappa _{2}\sin ^{2}\theta (\tau)\right) d\tau,
\end{equation}
which describes the loss of fidelity via qubit decay. The integrands has a rather transparent physical meaning: it simply reflects the decay of the time-dependent dark state Eq.~(\ref{dark}). The next order contribution is:
\begin{equation}\label{F2}
F^{\left( 2\right) }\simeq -\frac{\dot{\theta}(0)^{2}}{G(0)^{2}}-\frac{\dot{%
\theta}(t_{\text{f}})^{2}}{G(t_{\text{f}})^{2}}+\frac{2\dot{\theta}(0)\dot{%
\theta}(t_{\text{f}})}{G(0)G(t_{\text{f}})}\cos \int_{0}^{t_{\text{f}%
}}d\tau G(\tau ) , 
\end{equation}%
which is due to the non-adiabatic correction (i.e., the leakage out of the instantaneous eigenstate $|\tilde{2}\left(
t\right) \rangle $). It vanishes when the system approaches the deep adiabatic limit $\dot{\theta}(t)\ll G(t) $. The last term has an oscillating dependence with respect to $t_{\rm f}$ due to the spin-1 precession mentioned earlier. Finally, the dissipation of the quantum bus only enters the 3rd order contribution
\begin{equation}\label{F3}
F^{(3)}\simeq -\gamma \int_{0}^{t_{\text{f}}}d\tau \text{\ }\frac{\dot{\theta%
}(\tau )^{2}}{G(\tau )^{2}}-\gamma t_{\text{f}}\frac{\dot{\theta}(0)\dot{%
\theta}(t_{\text{f}})}{2G(0)G(t_{\text{f}})}\cos\int_{0}^{t_{\text{f}%
}}d\tau G(\tau ), 
\end{equation}%
as the quantum bus can only be populated through second-order
non-adiabatic leakage [see Fig.~\ref{schematics}(c)]. This
dissipation effect can be suppressed by long operation time and, not surprisingly, shows the same
type of oscillating behavior of the non-adiabatic correction.

\emph{Optimizing the STIRAP operation.-} Based on the perturbative treatment, we investigate the upper bound of the transfer fidelity and the best optimization strategy. We start with a special choice of the coupling profiles known as \emph{parallel adiabatic passage} (PAP)~\cite{Dridi2009}, where
$G_{1}(t)=G_{0}\sin \theta (t)$ and $G_{2}\left( t\right) =G_{0}\cos \theta(t)$. PAP is characterized by a constant energy splitting $G(t)=G_0$ and is commonly adopted by STIRAP protocols, as it allows one to suppress leakage errors by avoiding anticrossing points (see, for
example, Refs.~\cite{Guerin2002,Vasilev2009}). Due to the equal maximum couplings, PAP is a natural choice in the case of identical or similar qubits, which motivates us to take $\kappa_1=\kappa_2=\kappa$. The extension to the asymmetric case will be discussed later.

\begin{figure}
\begin{centering}
\includegraphics[width=0.48 \textwidth]{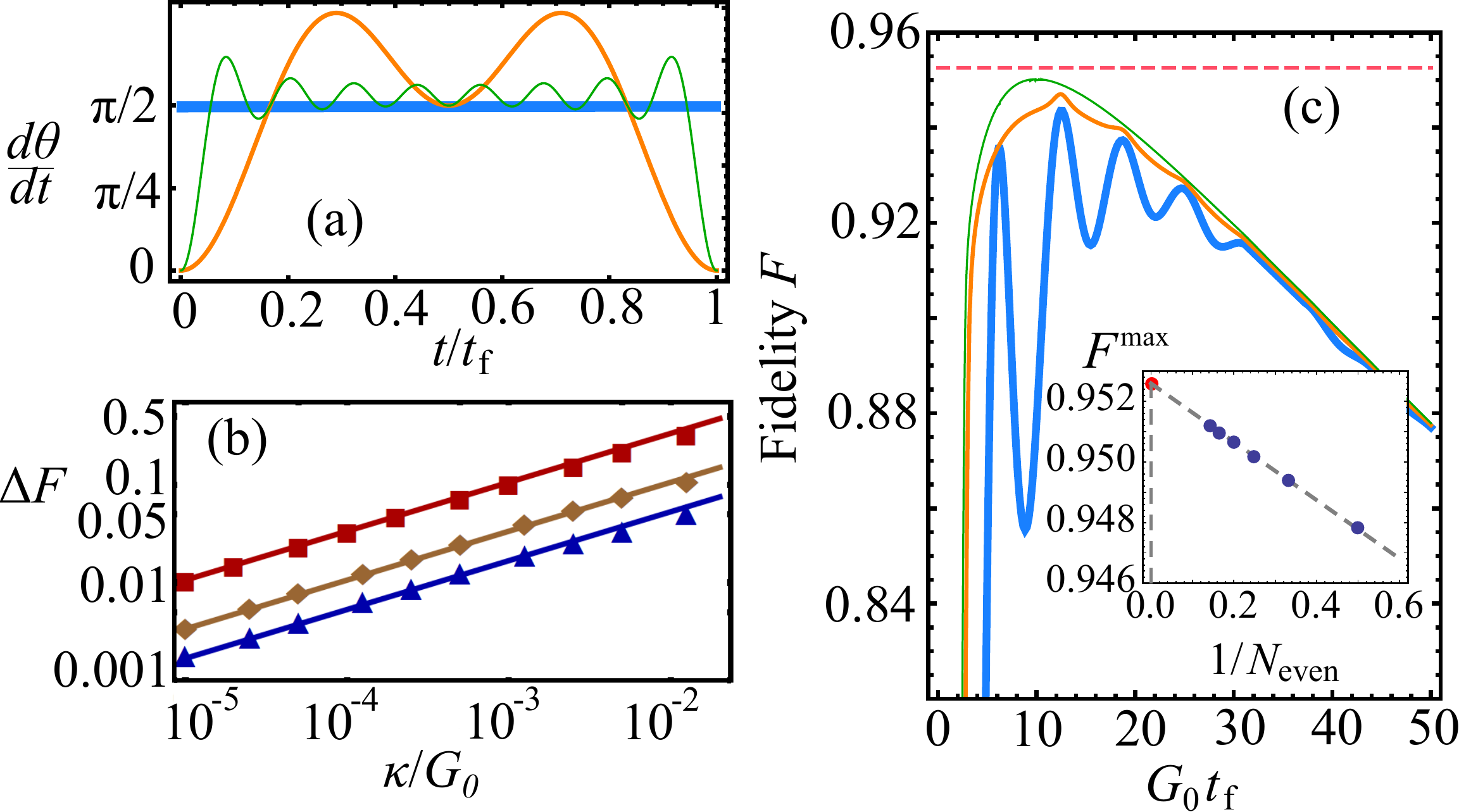} 
\par\end{centering}
\caption{Parallel adiabatic passage. (a) The optimized $\dot{\theta} (t)$ for $N_{\rm even}=0, 1, 4$ (line style from thick to thin). (b) The comparison of numerically optimized fidelity (data points) with the analytical result (solid lines). The numerical optimization is carried over $t_{\rm f}$ and $\alpha_{1,2,3,4}$ ($N_{\rm even}=2$) while the analytical result is $\Delta F=2\pi\sqrt{\xi(2)/C}$. The agreement is excellent, except small deviations at the larger values of $\kappa$ (as expected, due to the perturbative treatment). The three lines from bottom to top corresponds to $\gamma/G_0=0.025, 0.1, 1$. (c) Fidelity vs. transfer time, obtained by numerical optimization over $\alpha_n$ (cf. Eq.~(\ref{Fourier})). Solid curves are for $N_{\rm even}=0,1,4$ (bottom to top, also thick to thin). The horizontal dashed line indicates the maximum fidelity extrapolated for $N_{\rm even}\to \infty$ (see inset). Inset: Each blue dot are numerically optimized maximum fidelity for different $N_{\rm even}$ and the red dot marks the extrapolation to $N_{\rm even}\rightarrow \infty$. We have used $\kappa/G_0=2.5\times 10^{-3}$ and $\gamma/G_0=0.1$.  The improvement from a larger $N_{\rm even}$ is less significant than optimizing over $t_{\rm f}$.}
\label{fig:ana_num}
\end{figure}

For PAP, $F^{(1)}=-\kappa t_{\rm f}$ and the dependence on $\theta(t)$ vanishes. $\dot\theta(t)$  is the  only function to optimize and it can be written as a Fourier series: 
\begin{equation}
\dot{\theta}\left( t \right) =\alpha _{0}+\sum_{n>0}\alpha _{n}\cos
\left( n\pi t/t_{\mathrm{f}}\right) , \label{Fourier}
\end{equation}%
where $\alpha _{0}=\pi /(2t_{\mathrm{f}})$ is fixed by the boundary
condition Eq.~(\ref{boundary}) and $\alpha _{n}$ ($n>0$) represent a set of
optimization parameters. At a certain transfer time $t_{\mathrm{f}}$, optimizing $\dot{\theta}(t)$ requires $\partial F  /\partial \alpha _{n}=0$, which can be easily solved since the fidelity is a quadratic form of the $\alpha _{n}$. The  solution is:
\renewcommand{\arraystretch}{1.5}
\begin{equation}
 \alpha _{n }^{\mathrm{opt}} = \left\{
 \begin{array}{ll}
 \dfrac{-\pi / 2t_{\mathrm{f}}}{N_{\mathrm{even}}+\frac{\gamma t_{\mathrm{f}}}{4+(\gamma t_{\mathrm{f}}-4)\cos G_{0}t_{\mathrm{f}}}} ~~& {\rm for}~  n~{\rm even},\\
 0  & {\rm for}~ n~{\rm odd},\\
\end{array} \right.
 \label{alphan} 
\end{equation}%
i.e., to obtain the optimal fidelity, the coefficients of the even terms are all
equal; while the odd-$n$ Fourier components vanish, which is a consequence of the
symmetric setup ($\kappa _{1}=\kappa _{2}$). In Eq.~(\ref{alphan}), $N_{\mathrm{even}}$ is the
total number of even-$n$ ($n>0$) Fourier terms. Some examples of the resulting form of $\dot\theta(t)$ are shown in Fig.~\ref{fig:ana_num}(a).

\renewcommand{\arraystretch}{1}

Including more optimization parameters yields a higher fidelity. In the limit $N_{\mathrm{even}}\rightarrow \infty $, we get $\alpha _{n \in \mathrm{even}}^{\mathrm{opt}}\simeq -\pi /\left( 2t_{\mathrm{f}}N_{\mathrm{%
even}}\right) $ and the optimized fidelity over coupling profiles is $F^{%
\mathrm{opt}}(t_{\mathrm{f}})=1-\kappa t_{\mathrm{f}}-\pi ^{2}\gamma /\left(
4G_{0}^{2}t_{\mathrm{f}}\right) $, which shows the competition between qubit
decay and non-adiabatic transition with respect to the transfer time. If we further optimize
over $t_{\mathrm{f}}$, the maximum transfer fidelity is:
\begin{equation}
F^{\mathrm{max}} =1-2\pi/\sqrt{C},  \label{fmax}
\end{equation}
with the corresponding optimal transfer time $t_{\mathrm{f}}^{\mathrm{opt}}=\pi /( \kappa \sqrt{C}) $, and $C=4G_{0}^{2}/\gamma \kappa $ is the system cooperativity. This expression provides the largest attainable fidelity when an arbitrary state is transferred with a STIRAP-like pulse in a symmetric configuration and for a given set of parameters $G_{0}$, $\kappa $ and $\gamma $. Defining the loss of fidelity $\Delta F=1-F$, we show in Fig.~\ref{fig:ana_num}(b) that Eq.~(\ref{fmax}) is in excellent agreement with the numerically optimized fidelity, even with $\gamma \sim G_0$. Thus, Eq.~(\ref{fmax}) shows that the ultimate power of STIRAP is limited by the cooperativity whereas the transfer by sequential swapping requires the strong coupling condition (the fidelity to swap $|1\rangle \rightarrow |2\rangle$ followed by $|2\rangle \rightarrow |3\rangle$ is $F\sim 1- (\kappa+\gamma)/G_0$, thus requires $\gamma,\kappa \ll G_0$).

The maximum fidelity for a finite number of optimization parameters can also be found as 
$F^{\mathrm{max}}\left( N_{\mathrm{even}}\right) =1-2\pi \sqrt{\xi \left( N_{\mathrm{even}}\right) / C}$
with $\xi (N_{\mathrm{even}}) \equiv (N_{\mathrm{even}}+3/2)/(N_{\mathrm{even}}+1)$ \cite{EPAPS}. Thus, for a given setup, the goal of the fidelity (i.e., the acceptable deviation form the upper bound) determines $N_\mathrm{even}$, which subsequently determines the coupling profiles and operation time. Systematic improvements of the pulse shape can be gained by progressively increasing the number $N_{\rm even}$ of optimization parameters. 

It is interesting to notice that, even taking $N_{\mathrm{even}}=0$, i.e., adopting the simple dependence $\theta(t)=\pi t/(2 t_{\rm f})$, the transfer fidelity is $F^{\mathrm{max}}\left( N_{\mathrm{even}}=0\right) \approx 1-1.22({2\pi}/{\sqrt{C}}) $ at $t_{\mathrm{f}}^{\mathrm{opt}}\left( 0\right) \approx 1.22\pi /(\kappa \sqrt{C})$, which is just sightly smaller than $F^{\mathrm{max}}$. This observation suggests that, for a STIRAP-based state transfer under dissipation, optimizing the operation time (i.e., the operation speed) is far more efficient than introducing a complicated pulse. This is demonstrated in Fig.~\ref{fig:ana_num}(c), where the maximum fidelity grows quickly with $N_{\rm even}$ and moderate values are sufficient to achieve a small deviation from the upper bound.

\begin{figure}
\begin{centering}
\includegraphics[width=0.48\textwidth]{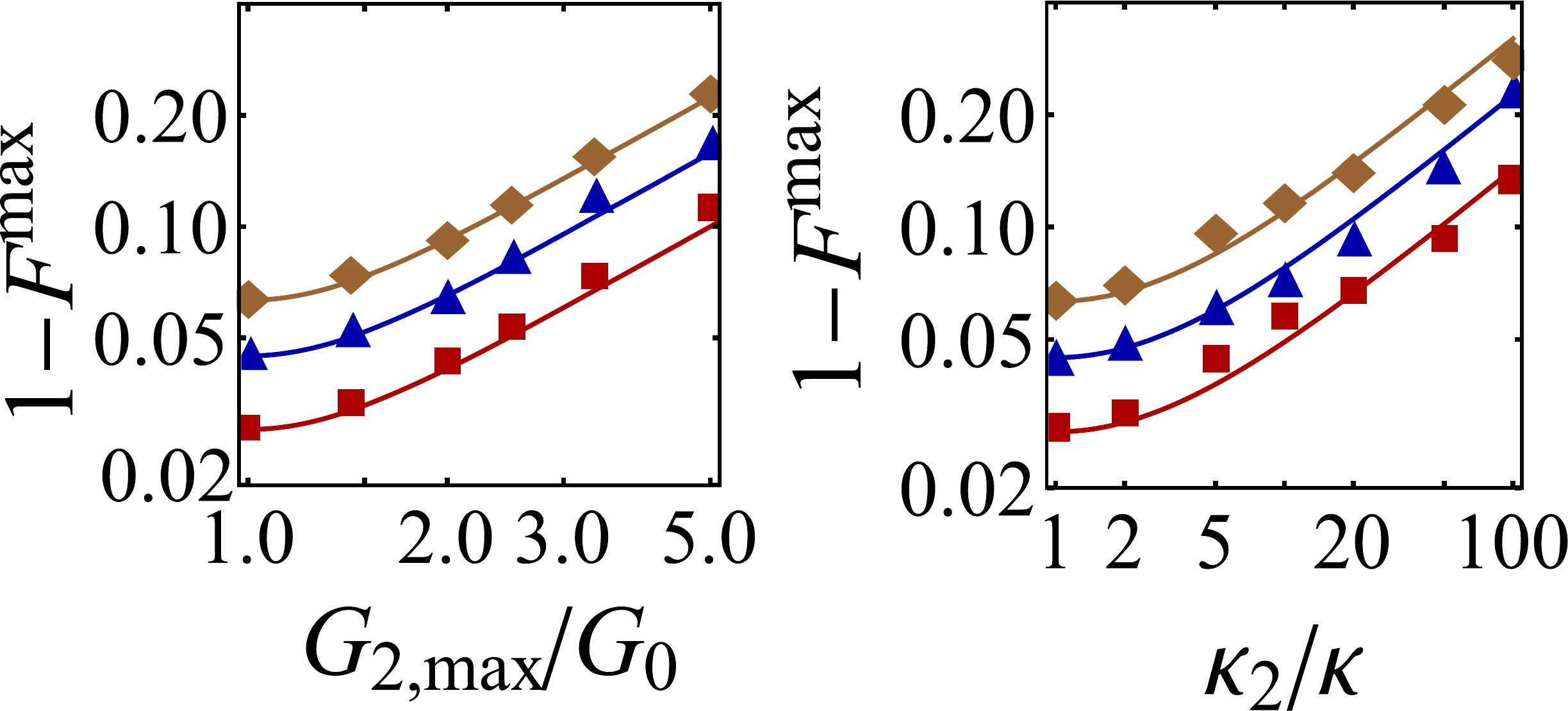} 
\par\end{centering}
\caption{Transfer beyond PAP. The markers are obtained from numerical optimizations, based on the coupling profiles $G_{i}(t)= G_{i,\rm max}\cos[\theta_i(t)-(2-i)\pi/2]$, with $i=1,2$. $\theta_i(t)$ is determined by optimization parameters $\alpha^{(i)}_{1,2,3,4}$, as defined in Eq.~(\ref{Fourier}). (a) Fidelity vs. $G_{2,\rm max}$ while $G_{1,\rm max}=G_0^2/G_{2,\rm max}$ and $\kappa_1=\kappa_2=\kappa$.  The three series of data are for $(\gamma/G_0,\kappa/G_0)=(0.2,0.025)$, $(0.1,0.025)$, and $(0.1,0.001)$ (top to bottom). The solid lines are from Eq.~(\ref{general}). (b) Asymmetric decay rates, satisfying $\kappa_1\kappa_2=\kappa^2$, with $(\gamma,\kappa)$ as in panel (a) and $G_{1,\rm max}=G_{2,\rm max}=G_0$. The solid lines are from Eq.~(\ref{Fmax_unequalk}). }
\label{fig_asymmetric_qubits}
\end{figure}

We now go beyond PAP by considering generic coupling profiles. Notice that (see Eq.~(\ref{F2}-\ref{F3})) the key variable in $F^{(2)}$ and $F^{(3)}$ is $\dot \theta(t)/G(t)$, which can be similarly decomposed in a Fourier series:
\begin{equation}
\dot \theta(t)/G(t)=\beta_{0}+\sum_{n>0}\beta_{n}\cos
\left(n\pi t/t_{\mathrm{f}}\right).
\end{equation}
The main difference is that $\beta_0 =t_{\rm f}^{-1}\int_{0}^{t_{\rm f}}d\tau \dot\theta(\tau)/G(\tau)$ cannot be fixed by the boundary condition, but it is constrained by the maximum coupling strengths $G_{1,\rm max}$ and $G_{2,\rm max}$~\cite{EPAPS}, so we can still perform the minimization with respect to $\beta_n$. The procedure closely patterns the PAP case, for example, $\beta^{\rm opt}_{n \in \rm even}$ is given by an expression similar to Eq.~(\ref{alphan}) except that the pre-factor $\pi/(2t_{\mathrm{f}})$ is replaced by $\beta_0$, and $\cos G_{0}t_{\mathrm{f}}$ by $\cos (\int_0^{t_{\rm f}}G(\tau)d\tau  )$. The optimization over the pulse shape yields ($N_{\rm even}\to \infty$)~\cite{EPAPS}:
\begin{eqnarray} \label{fmax2}
F^{\mathrm{opt}} (t_{\rm f})= 1-\kappa t_{\rm f}-\frac{\gamma}{t_{\rm f}}
\left( \frac{1}{G_{1,\rm max}^{2}}  +\frac{1}{G_{2, \rm max}^{2}}\right).
\end{eqnarray}%
Further optimizing over $t_{\rm f}$ leads to:
\begin{equation}
F^{\mathrm{max}} = 1-4\sqrt{\frac{1}{C_1}+\frac{1}{C_2}},  ~\qquad {\rm for~} \kappa_1=\kappa_2,\label{general}
\end{equation}
where we indicate with $C_i=4 G_{i,\rm max}^2/\gamma \kappa_i$ the cooperativity of side $i=1,2$. In the limit $C_1=C_2=C$, this result allows for a slightly larger fidelity than Eq.~(\ref{fmax}) ($4\sqrt{2}/(2\pi) \simeq 0.9$). This can be attributed to the fact that here $G(t)$ is not kept constant. Equation~(\ref{general}) also shows that in the limit when $C_1$ and $C_2$ are very different, the less coherent system (with lower $C_i$) dominates the fidelity loss. In Fig.~\ref{fig_asymmetric_qubits}(a) we compare our analytic expression to the numerical optimization, and show that Eq.~(\ref{general}) is indeed an accurate characterization of the maximum fidelity.

Finally, we consider a large difference in the qubit coherence (e.g., $\kappa_1 \gg \kappa_2$). Since $F^{(1)}$ also depends on $\theta(t)$ (see Eq.~(\ref{F1})), the previous analytic approach is not easily applicable. Still, one can understand the general parametric dependence by replacing the second term in Eq.~(\ref{fmax2}) with $F^{(1)}\sim -(\kappa_1 + \kappa_2)t_{\rm f}/2$. Note that large deviations from this estimate occur when $\theta(t)$ spends a significant fraction of the transfer time close to the initial or final angles $\theta=0,\pi/2$. However, this situation is essentially equivalent to a shorter $t_{\rm f}$. The argument indicates that the relevant figure of merit is the smallest cooperativity $C_{\rm min} =\sqrt{\gamma \kappa_{\rm max}}/G_{\rm min}^2$, where $\kappa_{\rm max}={\rm max}[\kappa_1,\kappa_2]$ and $G_{\rm min}={\rm min}[G_{1,\rm max},G_{2,\rm max}]$. We find that this conclusion is generally in agreement with the numerics. For example, Fig.~\ref{fig_asymmetric_qubits}(a) shows that when $G_{1,\rm max}=G_{2,\rm max}$, the maximum fidelity is well approximated by:
\begin{equation}\label{Fmax_unequalk}
F^{\rm max} \simeq 1- 2\left(\sqrt{\frac{2}{C_1}} +\sqrt{\frac{2}{C_2}}\right),  \quad {\rm for~} G_{1,{\rm max}}= G_{2,{\rm max}},
\end{equation}
which is in agreement with Eq.~(\ref{general}) when $C_1=C_2$. When $C_1$ is very different from $C_2$, Eqs.~(\ref{general}) and (\ref{Fmax_unequalk}) differ in the numerical prefactors but confirm the general argument that $\Delta F \sim 1/\sqrt{C_{\rm min}}$, for an optimized transfer time $t_{\rm f}^{\rm opt}\sim 1/(\kappa_{\max}\sqrt{C_{\rm min}})$.

\emph{Conclusion and Remarks.-} 
We have analyzed the STIRAP-based state transfer between two qubits interacting with a common harmonic mode. The trade-off between the non-adiabatic transitions and dissipation needs careful optimization of
the time-dependent couplings. Instead of a numerical approach, we have developed an analytical
treatment on the general parameter dependence, which has allowed us to reach a physical understanding of
the optimal transfer time and the upper bound of the fidelity. We also showed how to reach this bound efficiently. Our results provide a useful guideline for implementations of such protocol in a variety of physical systems, as well as
generalizations to alternative setups and more complicated adiabatic transfer schemes.

YDW and SC acknowledge the support from Chinese Youth 1000 Talents Program.
YDW also acknowledges the NSFC grants (No. 11574330 and No. 11434011). SC acknowledges the NSFC grant (No. 11574025). XBY is supported by China Postdoctoral Science Foundation (Grant No.
2015M571136). YDW also thank A. A. Clerk and H. Ribeiro for discussions at the early stage of this work.

\bibliography{stirap}

\pagebreak
\widetext
\begin{center}
\textbf{\large Supplemental material for ``Optimization of STIRAP-based state transfer under dissipation"}
\end{center}
\setcounter{equation}{0}
\setcounter{figure}{0}
\setcounter{table}{0}
\setcounter{page}{1}
\makeatletter
\renewcommand{\theequation}{S\arabic{equation}}
\renewcommand{\thefigure}{S\arabic{figure}}
\renewcommand{\bibnumfmt}[1]{[S#1]}
\renewcommand{\citenumfont}[1]{S#1}

\section{Unitary dynamics and transfer fidelity}

We discuss briefly here a simple optimization of the fidelity based on an exact solution of the unitary dynamics, and compare the STIRAP scheme with other types of state-transfer protocols.

First we consider Eq.~(3) of the main text and note  that the first two terms on the right side yield an
effective Hamiltonian 
\begin{equation}
\tilde{H}+i\frac{\dot{\theta}(t)}{\sqrt{2}}\mu =G(t)J_{z}+\dot{\theta}%
(t)J_{y}\equiv \vec{M}(t)\cdot \vec{J},  \label{Hmagn}
\end{equation}%
where we map $|\tilde{1}\rangle ,|\tilde{2}\rangle ,|\tilde{3}\rangle $ to the spin-1 angular
momentum states ($|+1\rangle ,|0\rangle ,|-1\rangle $, respectively). As seen,
the time evoultion is equivalent to the rotation of a spin in a time-dependent
magnetic field (see Fig.~1(d) in the main text).  For a coupling profile of the simple sin-cos form:
\begin{equation}
G_{1}(t)=G_{0}\sin (\pi t/2t_{\mathrm{f}}),\quad G_{2}(t)=G_{0}\cos (\pi
t/2t_{\mathrm{f}}),  \label{sinpulse}
\end{equation}
the magnetic field is constant and tilted from the $z$ direction by an angle $\phi =\arctan[\pi /(2G_{0}t_{\mathrm{f}})]$. Since the initial state is an eigenstate of $J_{z}$ (i.e., is not aligned to $\vec{M}$), an oscillatory dynamics follows, with angular frequency given by $|\vec{M}|=\sqrt{G_{0}^{2}+(\pi /2t_{\mathrm{f}})^{2}}$. The exact solution for the fidelity is: 
\begin{equation}
F=\left( \frac{(G_{0}t_{\mathrm{f}})^{2}+(\pi /2)^{2}\cos \sqrt{(G_{0}t_{%
\mathrm{f}})^{2}+(\pi /2)^{2}}}{(G_{0}t_{\mathrm{f}})^{2}+(\pi /2)^{2}}%
\right) ^{2}.  \label{Fsin}
\end{equation}%
Besides recovering $F\rightarrow 1$ in the adiabatic limit $t_{\mathrm{f}%
}\gg 1/G_{0}$, we also notice that, even for a much shorter operation time $%
t_{\mathrm{f}}\sim 1/G_{0}$, a perfect transfer is possible at the discrete
values $G_{0}t_{\mathrm{f}}=2\pi \sqrt{n^{2}-(1/4)^{2}}$ ($n=1,2,\ldots $).
Such a condition can be matched by tuning the transfer time $t_{%
\mathrm{f}}$ or the coupling strength $G_{0}$. Physically, these times
correspond to the periodic return of the state to the $z$ direction.

It is certainly not surprising that an ideal state transfer can be realized
in the absence of dissipation. In fact, more efficient and natural ways
exist. Among these, we would like to mention the well-known sequential swapping,
where a constant value of $G_{1}$ (with $G_{2}=0$) transfers $\left\vert 1\right\rangle \rightarrow \left\vert
2\right\rangle $ and then a second swap leads to $\left\vert 2\right\rangle
\rightarrow \left\vert 3\right\rangle $. The total transfer time is $\pi
/G_{0}$ (if both coupling strengths are equal: $G_1=G_2=G_0$), which is slightly shorter than the
minimum transfer time $2\pi \sqrt{1-(\pi /4)^{2}}/G_{0}\simeq 1.2\pi /G_{0}$
implied by Eq.~(\ref{Fsin}). Another way which is operationally simplest is
to set both couplings to a constant $G_{0}$ and let the system evolve for a
time $\pi /\sqrt{2}G_{0}$, which in the spin-1 language corresponds to a $%
\pi $-rotation around the $x$ axis. 

Nevertheless, the choice Eq.~(\ref{sinpulse}) is interesting in the context of the present discussion. On one hand, it shows that a STIRAP-like time-dependence of the couplings still allows for a
perfect state transfer deep in the non-adiabatic regime (i.e., with $t_{
\mathrm{f}}\sim 1/G_{0}$). On the other hand, this exact solution can serve
as a useful reference for our discussion of the general case, including
dissipation and more complicated choice for $\theta (t)$. In fact, we can use the solution  Eq.~(\ref{Fsin}) to estimate the fidelity loss due to the decay of the intermediate system: Without dissipation, the population of the excited quantum bus due to the unitary time evolution is:
\begin{equation}
\langle 2|\rho ^{I}\left( t\right) |2\rangle =\left[ \frac{\pi G_{0}t_{%
\mathrm{f}}}{2}\frac{1-\cos \left( t\sqrt{G_{0}^{2}+(\pi /2t_{\mathrm{f}%
})^{2}}\right) }{(G_{0}t_{\mathrm{f}})^{2}+(\pi /2)^{2}}\right] ^{2}.
\end{equation}%
The population loss through the bus is approximately:
\begin{equation}
\gamma
\int_{0}^{t_{\mathrm{f}}}\langle 2|\rho ^{I}|2\rangle dt\simeq \frac{3\pi
^{2}\gamma}{ 8G_{0}^{2}t_{\mathrm{f}} }.
\end{equation}
As discussed in the main text, this result is in agreement with the perturbative calculation. Combining this result with the cavity damping $\kappa t_{\mathrm{f}}$, one can optimize the transfer time and find the fidelity $F^{\rm max}(N_{\rm even}=0)=1-1.22\frac{2\pi}{\sqrt{C}}$. Thus, this simple example allows one to understand the relevant physics and typical time scales characterizing the state transfer process, while further optimization of $\theta (t)$ leads to a modest improvement of the fidelity.

\section{Perturbative solution of the master equation}

We present here the details of the perturbative treatment, which are too cumbersome to include in the main text. 

The  perturbative expansion is most conveniently carried out in the instantaneous eigenbasis
 $\left\{ |\tilde{k}(t)\rangle \right\} $, which is related to the time-independent basis  $\left\{| k(t)\rangle \right\} $ as follows:
\begin{eqnarray}
\left\vert \tilde{1}(t)\right\rangle  &=&\frac{1}{\sqrt{2}}\left( \sin
\theta (t)\left\vert 1\right\rangle +\left\vert 2\right\rangle +\cos \theta
(t)\left\vert 3\right\rangle \right) ,  \notag \\
\left\vert \tilde{2}(t)\right\rangle  &=&-\cos \theta (t)\left\vert
1\right\rangle +\sin \theta (t)\left\vert 3\right\rangle ,  \notag \\
\left\vert \tilde{3}(t)\right\rangle  &=&\frac{1}{\sqrt{2}}\left( -\sin
\theta (t)\left\vert 1\right\rangle +\left\vert 2\right\rangle -\cos \theta
(t)\left\vert 3\right\rangle \right) ,
\end{eqnarray}%
while $|\tilde{4}(t)\rangle =|4\rangle $. As defined in the main text, the original 4-level subspace is $\left\vert 1\right\rangle =\left\vert e^{\left( 1\right) },0,0^{\left(
2\right) }\right\rangle$, $\left\vert 2\right\rangle =\left\vert
g^{\left( 1\right) },1,0^{\left( 2\right) }\right\rangle$, $\left\vert 3\right\rangle =\left\vert g^{\left( 1\right) },0,1^{\left(
2\right) }\right\rangle$, $\left\vert 4\right\rangle =\left\vert
g^{\left( 1\right) },0,0^{\left( 2\right) }\right\rangle$, while the angle $\theta (t)$ is given by $\tan \theta (t)=G_{1}(t)/G_{2}(t)$. 

In the reference frame defined by  $\left\{ |\tilde{k}(t)\rangle \right\} $, we write the density matrix as $\tilde{\rho}(t)=\tilde{\rho}^{(0)}(t)+
\tilde{\rho}^{(1)}(t)+\tilde{\rho}^{(2)}(t)+\cdots $, where the lowest-order satisfies: 
\begin{equation}
\frac{d}{dt}\tilde{\rho}^{(0)}(t)=-i\left[ \tilde{H}(t),\tilde{\rho}^{(0)}(t)%
\right] .  \label{Order0}
\end{equation}
The solution is  simply
$\tilde{\rho}_{ab}^{\left( 0\right) }\left(t \right) =\delta_{a,2}\delta _{b,2}$, if the qubit 1 is initially in the excited state. The higher-order contributions satisfy: 
\begin{equation}
\frac{d\tilde{\rho}^{(k)}(t)}{dt}=-i\left[ \tilde{H}(t),\tilde{\rho}^{(k)}(t)%
\right] +\frac{\dot{\theta}(t)}{\sqrt{2}}\left[ \mu ,\tilde{\rho}^{(k-1)}(t)%
\right] +\mathcal{\tilde{L}}\tilde{\rho}^{(k-1)}(t), \label{rhokeq}
\end{equation}%
where the non-adiabatic term in matrix form is (for simplicity, we omit the explicit time dependence, i.e. $\tilde{\rho}=\tilde{\rho}(t)$)
\begin{equation}
\frac{\dot{\theta}}{\sqrt{2}}\left[ \mu ,\mathcal{\tilde{\rho}}\right] 
\mathcal{=}\dot{\theta}\left( 
\begin{array}{cccc}
\sqrt{2}{\rm Re}\mathcal{\tilde{\rho}}_{12} & \frac{\mathcal{\tilde{\rho}}%
_{22}-\mathcal{\tilde{\rho}}_{11}+\mathcal{\tilde{\rho}}_{13}}{\sqrt{2}} & 
\frac{\mathcal{\tilde{\rho}}_{23}-\mathcal{\tilde{\rho}}_{12}}{\sqrt{2}} & 0
\\ 
\frac{\mathcal{\tilde{\rho}}_{22}-\mathcal{\tilde{\rho}}_{11}+\mathcal{%
\tilde{\rho}}_{31}}{\sqrt{2}} & \sqrt{2}{\rm Re}\left( \mathcal{\tilde{\rho}%
}_{23}-\mathcal{\tilde{\rho}}_{12}\right)  & \frac{\mathcal{\tilde{\rho}}%
_{33}-\mathcal{\tilde{\rho}}_{22}-\mathcal{\tilde{\rho}}_{13}}{\sqrt{2}} & 0
\\ 
\frac{\mathcal{\tilde{\rho}}_{32}-\mathcal{\tilde{\rho}}_{21}}{\sqrt{2}} & 
\frac{\mathcal{\tilde{\rho}}_{33}-\mathcal{\tilde{\rho}}_{22}-\mathcal{%
\tilde{\rho}}_{31}}{\sqrt{2}} & -\sqrt{2}{\rm Re}\mathcal{\tilde{\rho}}_{23}
& 0 \\ 
0 & 0 & 0 & 0%
\end{array}%
\right) . \label{nonadiabatic}
\end{equation}%
The full expression of the dissipator is too cumbersome to present here. We only show the contribution from the quantum bus:
\begin{equation}
\mathcal{\tilde{L}}_{m}\mathcal{\tilde{\rho}=-\gamma }\left( 
\begin{array}{cccc}
\frac{\mathcal{\tilde{\rho}}_{11}+{\rm Re}\mathcal{\tilde{\rho}}_{13}}{2} & 
\frac{\mathcal{\tilde{\rho}}_{12}+\mathcal{\tilde{\rho}}_{32}}{4} & \frac{%
\mathcal{\tilde{\rho}}_{11}+\mathcal{\tilde{\rho}}_{33}+2\mathcal{\tilde{\rho%
}}_{13}}{4} & 0 \\ 
\frac{\mathcal{\tilde{\rho}}_{21}+\mathcal{\tilde{\rho}}_{23}}{4} & 0 & 
\frac{\mathcal{\tilde{\rho}}_{21}+\mathcal{\tilde{\rho}}_{23}}{4} & 0 \\ 
\frac{\mathcal{\tilde{\rho}}_{11}+\mathcal{\tilde{\rho}}_{33}+2\mathcal{%
\tilde{\rho}}_{31}}{4} & \frac{\mathcal{\tilde{\rho}}_{12}+\mathcal{\tilde{%
\rho}}_{32}}{4} & \frac{\mathcal{\tilde{\rho}}_{33}+{\rm Re}\mathcal{\tilde{%
\rho}}_{13}}{2} & 0 \\ 
0 & 0 & 0 & -\frac{\mathcal{\tilde{\rho}}_{11}+\mathcal{\tilde{\rho}}_{33}}{2} - {\rm Re}\mathcal{\tilde{\rho}}_{13}%
\end{array}%
\right) ,  \label{dism}
\end{equation}
which is independent on $\mathcal{\tilde{\rho}}_{22}$. Since at zero order $\mathcal{\tilde{\rho}}_{ij}^{(0)}=\delta_{i,2}\delta_{j,2}$, Eq.~(\ref{dism}) shows that $\tilde{\rho}^{(1)}(t)$ has no contribution proportional to $\gamma $. 
To evaluate the first-order correction to the fidelity, the full expression of $\left( \mathcal{%
\tilde{L}\tilde{\rho}}\right)_{22}$ is necessary:
\begin{eqnarray}
\left( \mathcal{\tilde{L}\tilde{\rho}}\right) _{22} &=&\mathcal{-}%
\left( \kappa _{1}\cos ^{2}\theta +\kappa _{2}\sin ^{2}\theta +\bar{\gamma}%
_{\varphi }\sin ^{2}2\theta \right) \mathcal{\tilde{\rho}}_{22}
+\frac{\kappa _{1}-\kappa _{2}}{4\sqrt{2}}\left( \mathcal{\tilde{\rho}}%
_{12}-\mathcal{\tilde{\rho}}_{23}+\mathrm{c.c.}\right) \sin 2\theta   \notag
\\
&&-\frac{\bar{\gamma}_{\varphi }}{2\sqrt{2}}\left( \mathcal{\tilde{\rho}}%
_{12}-\mathcal{\tilde{\rho}}_{23}+\mathrm{c.c.}\right) \sin 4\theta  
+\frac{\bar{\gamma}_{\varphi }}{2}\left( \mathcal{\tilde{\rho}}_{11}+%
\mathcal{\tilde{\rho}}_{33}-\mathcal{\tilde{\rho}}_{13}-\mathcal{\tilde{\rho}%
}_{31}\right) \sin ^{2}2\theta ,  \label{rho22}
\end{eqnarray}%
where $\bar{\gamma}_{\varphi }=\left( \gamma _{\varphi }^{\left( 1\right)
}+\gamma _{\varphi }^{\left( 2\right) }\right) /2$.

Equation~(\ref{rhokeq}) can be solved iteratively, to yield the perturbative expansion of the transfer fidelity $F=\tilde{\rho}_{22}(t_{\rm f})$. In particular, integrating Eq.~(\ref{rhokeq}) gives:
\begin{equation}\label{rhokeq2}
\tilde{\rho}_{ab}^{(k)}(t)=\int_{0}^{t}  d\tau
e^{-i \int_\tau^t \Delta E_{ab}(t')d t'} \left( \frac{\dot{\theta}(t)}{\sqrt{2}}\xi _{ab}^{(k-1)}(\tau
)+L_{ab}^{(k-1)}(\tau )\right),
\end{equation}%
where $\xi ^{\left( k-1\right) }(\tau )\equiv \lbrack \mu \mathbf{\mathbf{,}}%
\tilde{\rho}^{(k-1)}(\tau )]$, $L_{ab}^{(k-1)}(\tau )\equiv (\mathcal{%
\tilde{L}}\tilde{\rho}^{\left( k-1\right) }(\tau ))_{ab}$, and $\Delta E_{ab}(t)=E_a(t)-E_b(t)$, with $E_a(t) = G(t)(\delta_{a,1}-\delta_{a,3})$. Equation~(\ref{rhokeq2}), together with Eq.~(\ref{rho22}), yields the 1st order correction:
\begin{equation}
F^{\left( 1\right) }\mathcal{=-}\int_{0}^{t_{\mathrm{f}}}\left( \kappa
_{1}\cos ^{2}\theta (t^{\prime })+\kappa _{2}\sin ^{2}\theta (t^{\prime })+%
\bar{\gamma}_{\varphi }\sin ^{2}2\theta (t^{\prime })\right) dt^{\prime }.
\end{equation}%
This represents a generalization of Eq.~(4) of the main text (where we assumed $\bar{\gamma}_{\varphi }=0$).

At 2nd-order the disspation of the bus does not contribute to the fidelity either, because $[\mathcal{\tilde{L}}_{m}\mathcal{\tilde{\rho}]}_{22}$ is identically
zero [see Eq.~(\ref{dism})]. This observation, together with the independence of $\mathcal{\tilde{L}}_{m}\tilde{\rho}$ on $\tilde{\rho}_{22}$, reflects the fact that $|\tilde{2}\rangle $ does not involve excitations of the bus (it is a dark state). Thus, the population $\tilde{\rho}_{22}$ is not directly affected by a finite $\gamma$ and $\tilde{\rho}_{22}$ has no influence on $\mathcal{\tilde{L}}_{m}\tilde{\rho}$. The second order correction can be written as:
\begin{equation}
F^{\left( 2\right) }\simeq -2\int_{0}^{t_{\text{f}}}dt^{\prime }\dot{\theta}%
(t^{\prime })\int_{0}^{t^{\prime }}dt^{\prime \prime }\dot{\theta}(t^{\prime
\prime })\cos \left[ \int_{t^{\prime \prime }}^{t^{\prime }}d\tau G(\tau )%
\right] ,
\end{equation}
which takes into account the corruption of fidelity due to purely non-adiabatic leakage. While a full calculation of the 2nd-order result should take into account the qubits dissipation, these corrections are neglected here. This is due to the fact that the qubit dissipation already enters the 1st-order result and higher order terms involving $\kappa_{i},\gamma_\varphi^{(i)}$ should be much smaller.

We then consider the 3rd order correction, which finally yields a contribution proportional to $\gamma$. As illustrated schematically by Fig.~1(c) of the main text, the loss of fidelity via the mechanical damping is through the non-adiabatic leakage. This is also clear from Eqs. (\ref{nonadiabatic})  and (\ref{dism}). Consider, for example, that at 1st order the non-adiabatic leakage leads to a finite value $\tilde{\rho}_{12}^{(1)}\propto \dot{\theta}$. At 2nd order, this correction gives a contribution to $\tilde{\rho}_{12}^{(2)}\propto \dot{\theta}\gamma $ (cf. Eq. (\ref{dism})). Finally, the non-adiabatic terms leads to a correction to $\mathcal{\tilde{\rho}}_{22}$ from $\mathcal{\tilde{\rho}}_{12}^{(2)}$ (cf. Eq. (\ref{nonadiabatic})) $\mathcal{\tilde{\rho}}_{22}^{(3)}\propto \dot{\theta}^{2}\gamma $. The full expression reads:
\begin{equation}
F^{\left( 3\right) }\simeq -\gamma \int_{0}^{t_{\text{f}}}dt_{1}\dot{\theta}%
(t_{1})\int_{0}^{t_{1}}dt_{2}\sin \left( \int_{t_{2}}^{t_{1}}d\tau G(\tau
)\right) \int_{0}^{t_{2}}dt_{3}\dot{\theta}(t_{3})\sin \left(
\int_{t_{3}}^{t_{2}}d\tau G(\tau )\right) .
\end{equation} 

The above formulas for $F^{\left(2\right) }$ and $F^{\left( 3\right) }$ can be simplified in the relevant case of a sufficiently large $G(\tau )$. In fact, they are given by integrals of the form $\int_{0}^{t}dt^{\prime}f(t^{\prime })\exp \left[ \pm i\int_{0}^{t^{\prime }}d\tau G(\tau )\right] $, where $f(t^{\prime })$ is a relatively smooth function while $\exp \left[\pm i\int_{0}^{t^{\prime }}d\tau G(\tau )\right] $ is a fast oscillating factor. By performing multiple integrations by parts, a systematic expansion of such integrals in powers of $G(t)^{-1}$ can be derived. A straightforward but tedious calculation yields the leading-order results, cited in Eqs.~(5) and (6) of the main text.

\section{Maximum fidelity with unequal couplings}
In the PAP case, the discussion is based on equal maximum couplings and equal damping rates for both qubits. Here we relax the constrain of equal maximum couplings and investigate again the upper bound of the fidelity. 

To do this, we write $\dot\theta(t)/G(t)$ into a Fourier expansions with $\beta_n$ the new set of optimization parameters (see Eq.~(10) in the main text). This makes the optimization very similar to the PAP case, except two difficulties. The first one is that not only $\dot\theta(t)/G(t)$, but also $\cos\int_0^{t_{\rm f}}G(\tau)d\tau$ is affected by the coupling profiles (while in the PAP case this gives a simple constant, $\cos G_0 t_{\rm f}$). The second complication is that $\dot{\theta}(t)/G(t)$ must satisfy more involved constrains than $\dot\theta(t)$. In fact, at each value of $\theta(t)$ the coupling $G$ has a maximum value $G_{\rm max}(\theta)$ [see Eq.~(\ref{Gmax}) below]. It is not immediately clear how the condition $G(t)\leq G_{\rm max}(\theta(t))$ (together with the old one, $\int_0^{t_{\rm f}}\dot\theta(\tau)d\tau = \pi/2$) can be simply written in terms of the $\beta_n$. To avoid these difficulties, we first perform the maximization by considering the $\beta_{n>0}$ and $\cos\int_0^{t_{\rm f}}G(\tau)d\tau$ as arbitrary parameters. This approach is useful because removing these constrains gives an upper bound to the fidelity, at fixed $\beta_0$. Furthermore, in performing the calculation, we will also see how the above two points can be resolved.

$F$ can be written as a quadratic form of the $\beta_{n}$ since
\begin{equation}\label{aux_formulas1}
\frac{\dot\theta(0)}{G(0)}=\beta_0+\sum_{n>0} \beta_n, \quad\quad \frac{\dot\theta(t_{\rm f})}{G(t_{\rm f})}=\beta_0+\sum_{n>0} (-1)^n \beta_n, \quad {\rm and} \quad 
\int_{0}^{t_{\mathrm{f}}}d\tau \frac{\dot{\theta}(\tau )^2}{G(\tau)^2}=\beta_0^2 t_{\rm f}+\frac{t_{\rm f}}{2}\sum_{n>0} \beta_n^2.
\end{equation}
For even $n>0$, the maximization gives:
\begin{equation}
\frac{\partial F}{\partial \beta_n} = -4 \left(\beta_0+\sum_{n=2,4 \ldots} \beta_n \right) 
\left[1-\left(1-\frac{\gamma t_{\rm f}}{4}\right)\cos\int_0^{t_{\rm f}}G(\tau)d\tau\right]-\gamma t_{\rm f} \beta_n =0,
\end{equation}
showing that the optimized value of $\beta_n$ is independent of $n$: $\beta_n=\beta^{\rm opt}_{\rm even}$ and $\sum_{n=2,4 \ldots} \beta_n =N_{\rm even}\beta^{\rm opt}_{\rm even}$. Thus the analog of Eq.~(8) of the main text can be obtained easily
\begin{equation}
\beta _{\mathrm{even}}^{\mathrm{opt}} =-\beta_0 \frac{%
4+(\gamma t_{\mathrm{f}}-4)\cos\int_0^{t_{\rm f}}G(\tau)d\tau}{(4N_{\mathrm{even}%
}+\gamma t_{\mathrm{f}})+N_{\mathrm{even}}(\gamma t_{\mathrm{f}}-4)\cos\int_0^{t_{\rm f}}G(\tau)d\tau}. \qquad \label{betan} 
\end{equation}%
In a similar way, one can show that $\beta_{\rm odd}^{\rm opt}=0$. The fidelity optimized over the coupling profiles reads:
\begin{equation}  \label{Fopt_tf}
F_{\mathrm{opt}}(\beta_0,t_{\mathrm{f}}) \leq   1-\kappa t_{\mathrm{f}}-\beta_0^2  \frac{\gamma  t_{\rm f}  }{2 N_{\mathrm{even}}} \bigg(2N_{\mathrm{%
even}}+1 -\frac{\gamma t_{\mathrm{f}}} {4N_{\mathrm{even}}+\gamma t_{\mathrm{f}%
}+N_{\mathrm{even}}(\gamma t_{\rm f}-4) \cos\int_0^{t_{\rm f}}G(\tau)d\tau} \bigg). \qquad
\end{equation}
Since $F_{\mathrm{opt}}(t_{\mathrm{f}})$ is (as expected) a monotonic function of $N_{\rm even}$, the largest value is obtained by taking the limit $N_{\rm even}\to \infty $. In this case, the factor $\cos\int_0^{t_{\rm f}}G(\tau)d\tau$ drops out of the final expressions:
\begin{equation}\label{Fsimple_beta0}
F_{\mathrm{opt}}(\beta_0, t_{\mathrm{f}}) \leq \, 1-\kappa t_{\mathrm{f}} - \beta_0^2\gamma t_{\rm f}  .
\end{equation}
Thus, the specific value of  $\cos\int_0^{t_{\rm f}}G(\tau)d\tau$  is not important for the upper bound and the first difficulty is resolved. In the limit $N_{\rm even}\to \infty $, we also have $\beta _{\mathrm{even}}^{\mathrm{opt}} \simeq -\beta_0/N_{\rm even}$ and
\begin{equation}\label{ft_def}
\frac{\dot{\theta}(t)}{G(t)}  \simeq  \beta_0 \left(1-\sum_{m=1}^{N_{\rm even}}\frac{\cos{(2m\pi t/t_{\rm f})}}{N_{\rm even}}  \right) \equiv \beta_0 f(t).
\end{equation}

To take into account the constrain on $G(t)$ and further optimize the fidelity, we find now the lower bound of $\beta_0$. Since $\dot\theta (t) = \beta_0 G(t) f(t)$, with $G(t)$ and $f(t)$ both positive, $\dot\theta(t)$ has a well-defined sign (i.e., the same sign of $\beta_0$). We take $\dot\theta(t)>0$ and use the definition of $\beta_0$:
\begin{equation}\label{up_integral}
\beta_0=\frac{1}{t_{\rm f}}\int_0^{t_{\rm f}} \frac{\dot{\theta}(\tau)}{G(\tau)}d\tau \geq \frac{1}{t_{\rm f}} \int_0^{t_{\rm f}}  \frac{\dot{\theta}(\tau)}{G_{\rm max}(\theta(\tau))}d\tau,
\end{equation}
where $G_{\rm max}(\theta)$ is the maximum achievable coupling at a given value of the angle $\theta=\arctan(G_1/G_2)$. If $\bar\theta$ is the angle with both couplings maximized (i.e., $\bar\theta=\arctan(G_{1,\rm max}/G_{2, \rm max})$), $G_{\rm max}(\theta) $ is given by:
\begin{equation}\label{Gmax}
G_{\rm max}(\theta) = \left \{
\begin{array}{ll}
\frac{G_{2, \rm max}}{\cos{\theta}} & {\rm if}~~ 0 \leq \theta\leq \bar\theta, \\
& \\
\frac{G_{1, \rm max}}{\sin{\theta}} & {\rm if}~~ \bar\theta < \theta\leq \frac{\pi}{2}, 
\end{array}
\right.
\end{equation}
which allows us to rewrite Eq.~(\ref{up_integral}) as:
\begin{equation}\label{ub2}
\beta_0 \geq \frac{1}{t_{\rm f}}\int_0^{\bar{t}}  \frac{\cos\theta(\tau)\dot{\theta}(\tau)}{G_{2,\rm max}}d\tau+\frac{1}{t_{\rm f}}\int_{\bar{t}}^{t_{\rm f}}  \frac{\sin\theta(\tau)\dot{\theta}(\tau)}{G_{1,\rm max}}d\tau.
\end{equation}
We supposed here that there is a single solution of $\theta(\bar{t})=\bar{\theta}$, but the argument is easily adapted to multiple solutions. The integration of Eq.~(\ref{ub2}) is immediate and, using the boundary conditions $\theta(0)=0$, $\theta(t_{\rm f})=\pi/2$, as well as elementary trigonometric relations to express $\sin\bar\theta$, $\cos\bar\theta$ in terms of the $G_{i,\rm max}$, we get:
\begin{equation}
\beta_0 \geq   \frac{\sin\bar\theta}{t_{\rm f}G_{2,\rm max}}+\frac{\cos\bar\theta}{t_{\rm f}G_{1,\rm max}} = \frac{1}{t_{\rm f}}\sqrt{\frac{1}{G_{1,\rm max}^2}+\frac{1}{G_{2,\rm max}^2}} .
\end{equation}
Using this inequality in Eq.~(\ref{Fsimple_beta0}) gives:
\begin{equation}\label{Fsimple_final}
F_{\mathrm{opt}}(t_{\mathrm{f}}) \leq \, 1-\kappa t_{\mathrm{f}} - \frac{ \gamma }{t_{\rm f}}\left(\frac{1}{G_{1,\rm max}^2}+\frac{1}{G_{2,\rm max}^2}\right) ,
\end{equation}
which is the Eq.~(11) of the main text (taking the equality sign).

It is also interesting to consider in more detail the properties of $\dot\theta(t)/G(t)$, which allows one to understand better how the minimization is achieved, and leads to a slightly different (and perhaps more transparent) derivation of Eq.~(\ref{Fsimple_final}).  First we notice, using Eq.~(\ref{ft_def}):
\begin{equation}\label{optimumpulse1}
\frac{\dot\theta(0)}{G(0)}=\frac{\dot\theta(t_{\rm f})}{G(t_{\rm f})} \simeq \beta_0 \left(  1- \sum_{m =1 }^{N_{\rm even}}\frac{1}{N_{\rm even}}  \right)= 0,
\end{equation}
showing that $F^{(2)}$ (the non-adiabetic contribution) vanishes for the optimal coupling profiles (see Eq. (5) of the main text). By taking $\dot\theta(0)=\dot\theta(t_{\rm f})=0$, the general formula for $F$ is simplified to:
\begin{equation}\label{F00}
\left. F \right |_{\dot\theta(0)=\dot\theta(t_{\rm f})=0}=1-\kappa t_{\rm f} -\gamma \int_0^{t_{\rm f}} d\tau \frac{\dot\theta(\tau)^2}{G(\tau)^2}.
\end{equation}
If we consider intermediate times $\Delta t < t < t_{\rm f}-\Delta t$ (with $\Delta t \sim t_{\rm f}/N_{\rm even}$), it is easy to see that the summation in Eq.~(\ref{ft_def}) gives a small value and $\dot\theta(t)/G(t) \simeq \beta_0$. Using this constant in the integral of Eq.~(\ref{F00}), $F_{\mathrm{opt}}(\beta_0, t_{\mathrm{f}})$ of Eq.~(\ref{Fsimple_beta0}) is immediately recovered.  

This analysis arrives at a  simple charactrerization of the optimum pulse: we should choose $\theta(t)$  to satisfy $\dot\theta(0)=\dot\theta(t_{\rm f})=0$ and, for intermediate times, try to minimize the integral in Eq.~(\ref{F00}). Thus, we can rederive Eq.~(\ref{Fsimple_final}) by relying directly on the minimizion of Eq.~(\ref{F00}), and without using the Fourier decompostion. Clearly, we have
\begin{equation}\label{integral1}
\int_0^{t_{\rm f}} d\tau \frac{\dot\theta(\tau)^2}{G(\tau)^2} \geq  \int_0^{t_{\rm f}} d\tau \frac{\dot\theta(\tau)^2}{G_{\rm max}(\theta(\tau))^2},
\end{equation}
which, using  Eq.~(\ref{Gmax}) is written:
\begin{equation}\label{integral2}
\frac{1}{G_{2,\rm max}^2}\int_0^{\bar{t}}  \left(\frac{d\sin\theta(\tau)}{d\tau}\right)^2 d\tau+
\frac{1}{G_{1,\rm max}^2} \int_{\bar{t}}^{t_{\rm f}}   \left(\frac{d\cos\theta(\tau)}{d\tau}\right)^2 d\tau.
\end{equation}
The two integrals are minimized when their integrands are constant, i.e., we can set $d\sin\theta(\tau)/d\tau = (\sin\bar\theta)/\bar t $ and $d\cos\theta(\tau)/d\tau = (\cos\bar\theta)/(\bar t -t_{\rm f}) $. We conclude that Eq.~(\ref{integral2}) is larger or equal to:
\begin{equation}\label{integral3}
\frac{1}{\bar t}\left(\frac{\sin\bar\theta}{G_{2,\rm max}}\right)^2+ \frac{1}{t_{\rm f}-\bar t}\left(\frac{\cos\bar\theta}{G_{1, \rm max}}\right)^2 \geq  \frac{1}{t_{\rm f}}\left(\frac{1}{G_{1,\rm max}^2}+\frac{1}{G_{2,\rm max}^2}\right),
\end{equation}
where in the last step we used the definition of $\bar \theta$ and performed the minimization with respect to $\bar t$, giving $\bar t = t_{\rm f} (\sin\bar\theta)^2$. Equation~(\ref{integral3}) is the desired result, in agreement with Eq.~(\ref{Fsimple_final}). This derivation also shows explicitly that it is possible to find a suitable time-dependence of $\dot\theta(t)/G(t)$ approaching the equality sign in Eq.~(\ref{Fsimple_final}).

\end{document}